\documentclass[%
 preprint,
superscriptaddress,
 amsmath,amssymb,
 aps,
]{revtex4-2}

\usepackage{graphicx}% Include figure files
\usepackage{dcolumn}% Align table columns on decimal point
\usepackage{bm}% bold math
\usepackage{subcaption}
\usepackage{float}
\usepackage{hyperref} 
\usepackage[english]{babel}
\usepackage{bbold}
\begin{document}
\renewcommand{\thefootnote}{\fnsymbol{footnote}}
\setcounter{footnote}{0}
%\preprint{APS/123-QED}

\title{A T-matrix database to promote information-driven research in nanophotonics}

\author{Nigar Asadova}
\thanks{These authors contributed equally to this work.
Corresponding author: nigar.asadova@kit.edu
}
\affiliation{Institute of Nanotechnology, Karlsruhe Institute of Technology, Kaiserstrasse 12, 76131 Karlsruhe, Germany}

\author{Kaoutar Boussaoud}
\thanks{These authors contributed equally to this work.
Corresponding author: nigar.asadova@kit.edu
}

\affiliation{Scientific Computing Center, Karlsruhe Institute of Technology, Kaiserstrasse 12, 76131 Karlsruhe, Germany}

\author{J\"org Meyer}
\affiliation{Scientific Computing Center, Karlsruhe Institute of Technology, Kaiserstrasse 12, 76131 Karlsruhe, Germany}

\author{Frank Tristram}
\affiliation{3D Matter Made to Order, Karlsruhe Institute of Technology, Kaiserstrasse 12, 76131 Karlsruhe, Germany}

\author{Carsten Rockstuhl}
\affiliation{Institute of Nanotechnology, Karlsruhe Institute of Technology, Kaiserstrasse 12, 76131 Karlsruhe, Germany}
\affiliation{Institute of Theoretical Solid State Physics, Karlsruhe Institute of Technology, Kaiserstrasse 12, 76131 Karlsruhe, Germany}
\affiliation{Center for Integrated Quantum Science and Technology (IQST), Karlsruhe Institute of Technology, Wolfgang-Gaede-Str. 1, 76131 Karlsruhe, Germany}

%\noindent\textsuperscript{$\dagger$}These authors contributed equally to this work.

 %% email address is required; see note below about the corresponding author designation

% use {asbstract*} to suppress the copyright line. Copyright information will be added in production

\begin{abstract} 
Information-driven methods from machine learning and artificial intelligence for exploring the optical response of metasurfaces and, more generally, photonic systems rely on well-annotated datasets for training. For metasurfaces made from a periodic or aperiodic arrangement of scatterers, the primary information encoding their response is the optical properties of these individual scatterers. In the linear regime, that response is entirely contained in the transition or T-matrix of the individual scatterer. However, despite the widespread use of these T-matrices in exploring advanced photonic materials within the larger community, there is no common infrastructure for distributing them with consistent metadata and a standard representation. That would be important to avoid the repetitive, resource-intensive computation of these T-matrices by researchers worldwide and to enable data-driven research. To overcome this limitation, we introduce the Daphona T-matrix portal at \url{https://tmatrix.scc.kit.edu/}, a web-based platform for interactive searching, filtering, and exporting standardized data containing structure-property relations for a wide range of scatterers, as expressed by their T-matrices. Besides introducing this infrastructure, we demonstrate how the available data enables addressing scientific questions in the broader context of information-driven research. The multiple illustrative examples in our contribution cover both surrogate forward models and inverse design models, and operate either directly on the T-matrix or alternatively on optical observables of metasurfaces made from these scatterers. 

\end{abstract}

% \begingroup
% \renewcommand{\thefootnote}{\fnsymbol{footnote}}
% \setcounter{footnote}{0}

% \begingroup
% \renewcommand{\thefootnote}{\fnsymbol{footnote}}
% \setcounter{footnote}{0}

\maketitle

% \footnotetext[1]{Corresponding author: nigar.asadova@kit.edu}
% \endgroup
% \begingroup
% \renewcommand{\thefootnote}{\fnsymbol{footnote}}
% \footnotetext[1]{\hypertarget{corremail}{}Corresponding author: \href{mailto:nigar.asadova@kit.edu}{nigar.asadova@kit.edu}}
% \footnotetext[2]{\hypertarget{equalcontrib}{}These authors contributed equally to this work.}
% \endgroup

%%%%%%%%%%%%%%%%%%%%%%%%%%  body  %%%%%%%%%%%%%%%%%%%%%%%%%%
\section*{Introduction}

The transition matrix (T-matrix) is a key mathematical tool for characterizing how light interacts with a spatially localized object. When an electromagnetic wave encounters an object, such as a nanoparticle, molecule, or any microstructure, it induces complex physical phenomena: part of the light is redirected, part is absorbed, and part excites multipolar resonances within the object. The T-matrix captures this complete scattering behavior in a compact form by expressing how the multipole components of an incident field are transformed into those of the scattered field. The key benefit is that by considering the T-matrix, we reformulate the solution to Maxwell's equations as an algebraic problem. A basic matrix-vector product expresses how an arbitrary illumination interacts with a given scatterer. And the best thing is that this matrix is not even large. Depending on the multipolar order retained in the expansion of the incident and scattered field, we speak of matrices at most with a size hundred elements squared.   
While the description of light scattering by an individual object can already be conveniently expressed, the T-matrix approach unfolds its strength when applied to a wide variety of modern photonic materials composed of many individual scatterers. Examples of such photonic materials that can be studied efficiently include metasurfaces, metamaterials, photonic crystals, colloidal assemblies, disordered materials, optical antennas, and many more~\cite{antosiewicz2014plasmonic, theobald2021simulation, pourjamal2019lasing, zhuromskyy2016applicability, yazhgur2021light,  zhang2001multiple}. When the T-matrices are computed using quantum-chemical methods for individual molecules, many molecular materials can be described very efficiently as well~\cite{zerulla2022multi, doi:10.1021/acs.jpcc.5c03310}. Moreover, while we restrict ourselves here to linear interactions, the T-matrix framework can be extended to accommodate nonlinear effects, thereby providing a comprehensive approach to studying light-matter interactions~\cite{wunderlich_molecular_2011, sekulic_t_2025, zerulla2024multi}. Because of its versatility, the T-matrix has become an essential concept across nano-optics, photonics, atmospheric science, and computational electromagnetics~\cite{mishchenko2002scattering,  wriedt2009light, gantzounis2006layer,  markkanen2017fast, hohenester_unified_2025, bi2014accurate, rahimzadegan2022comprehensive}. In particular, the atmospheric science and remote sensing community has been especially prolific in establishing scattering property databases for specific applications~\cite{Yang:05, MENG2010501, amt-13-6933-2020, cairo2023study}.

However, obtaining accurate T-matrices requires solving Maxwell’s equations under many illumination conditions, a process that is computationally intensive, time-consuming, and energy-demanding. This holds particularly true when exploring parameter sweeps involving geometry, orientation, material dispersion, and spectral ranges. A persistent challenge in the field is that T-matrices are frequently recalculated rather than systematically reused. Research groups often rely on solver-specific scripts and internal storage conventions, leading to a wide variety of inconsistent and mutually incompatible dataset formats~\cite{hellmers2010scattport, mishchenko2004t}. This variation in formats makes reproducibility and data sharing difficult and prevents the development of shared databases and repositories. Having such repositories would be beneficial from at least three perspectives:

First, the repeated computation of identical or similar structures not only wastes scientific resources but also increases unnecessary energy consumption, which conflicts with the growing emphasis on sustainable high-performance computing practices. Second, collecting, archiving, and sharing T-matrices would enable the application of a wide range of contemporary tools from the fields of artificial intelligence and machine learning to address novel, data-driven scientific research questions. Besides developing surrogate solvers for Maxwell's equations, all kinds of inverse problems could be addressed using artificial neural networks when trained only on sufficiently large data sets. And third, archiving T-matrices and making them accessible to everyone would allow others to reproduce simulations more easily. Thanks to the extensive ecosystem of tools that can accommodate and handle these T-matrices to study specific photonic materials, access to T-matrices supports our responsibility to make our research reproducible. There may be many more motivations for establishing a repository dedicated to T-matrices, but these are the most urgent.  

To overcome these limitations, recent efforts, including our own to define a unified T-matrix data format coordinated with a broader part of the community \cite{ASADOVA2025109310}, have highlighted the need for standardized files and well-structured metadata to make T-matrix datasets clear, compatible, and easy to reuse. The proposed HDF5~\cite{hdf5_2022} based format places the T-matrix, the definition of its electromagnetic modes, the scatterer’s geometry and materials, and all information about how the data was computed into a single, organized, machine-readable structure. This standard is essential for sharing T-matrices on a larger scale and for supporting FAIR (Findable, Accessible, Interoperable, Reusable) data practices~\cite{wilkinson2016fair} in the nano-optics community.

Building on this work, we introduce here a dedicated web portal designed to streamline the storage, exploration, and reuse of T-matrices. It is available at \url{https://tmatrix.scc.kit.edu}. The portal allows researchers to import standardized HDF5 files through multiple upload modes, including individual files, compressed archives (ZIP), or folder imports, and automatically extracts, validates, and indexes their metadata. A user-friendly interface supports advanced search, filtering, and visualization capabilities based on geometry, material, wavelength, and other key attributes. In addition, the platform includes an open API that enables automated querying, uploading, and retrieval of T-matrices, facilitating seamless integration with external simulation pipelines and machine-learning workflows. By bringing T-matrix data together into one shared platform, it represents a major step toward a community-wide ecosystem for standardized optical scattering data and lays the groundwork for future large-scale T-matrix repositories and analysis tools.

However, it is not just the database that we intend to introduce here. Simultaneously, we wish to outline a few illustrative examples of how the data could be used for addressing information-driven research questions in the broader context of photonic materials composed of scattering structures. Basically, we can categorize the research questions along different dimensions. On the one hand, we can speak of forward vs. inverse problems. In the forward problem, we wish to compute the optical response from a given structure~\cite{wiecha2019deep, augenstein2023neural}, whereas in the inverse problem, we aim to identify the structure that produces the desired optical response~\cite{so2020deep, kang_large-scale_2024, wiecha2021deep, ma2025benchmarking}. Generally, such structure-property relations can be learned efficiently using artificial neural networks trained on data from such a database. On the other hand, we can distinguish between scenarios in which the T-matrix (or another representation of optical response) of an object is relevant~\cite{jing_deep_2023, rezaei_faster_2024, BI2024109057, so_simultaneous_2019, zhang_spherical-harmonic-based_2025} and those in which the optical response of a photonic material made from such scatterers is~\cite{li_machine_2024, so_revisiting_2023, gladyshev2023inverse}. Many efforts focus on physics-informed neural networks or alternative optimization approaches~\cite{ deng_physics-informed_2025, zhelyeznyakov_large_2023, zhelyeznyakov_design_2020, igoshin_inverse_2025, jackson_pymiediff_2025}. Within this contribution, we illustrate, with selected examples, how the data already available in the current database can be used to address such and similar research questions.

As an overarching idea and long-term vision, we propose developing an artificial neural network that serves as a general-purpose Maxwell solver for arbitrary structures across a wide range of frequencies. Once trained, inference can be performed with comparably low resources in a relatively short time, even for large-scale, spatially complex photonic materials. However, the training data for such a network should not be generated for the specific purpose of only generating the training data. That would consume unprecedented amounts of energy and resources, which would be hard to justify. However, if we, as a community, were to collect the outcomes from all our simulations, which we do anyway in the context of our daily research, into a common repository, we could harvest them for specific needs, particularly for the purpose of training a surrogate solver for the general Maxwell equations. While at the moment it is hard to think about how the outcome from the wide range of different Maxwell solvers can be combined, we do so here for the subset of those methods that rely in a broader sense on the scattering formalism.

The article is structured into four main sections. After the admittedly extended introduction that motivates the entire research, we briefly summarize in the second section the basic scientific background of the T-matrix formalism. In the third section, we outline the functionality and the access to the T-matrix database. In the fourth section, we demonstrate in multiple approaches how machine learning methods can be applied to this T-matrix data. The article ends with conclusions.

\section*{Summary on the T-Matrix approach}
The central idea behind the T-matrix approach to electromagnetic interaction is the expansion of electromagnetic fields at a given frequency in vector spherical waves (VSWs), regular for the incident field, and singular for the scattered field~\cite{waterman2005matrix}. These VSWs are analytical solutions to the frequency domain Maxwell's equations in a spherical coordinate system. The linear map between the amplitude coefficients of those VSWs expanding the incident field, represented by the vector $\mathbf{a}$, and the amplitude coefficients of those VSWs expanding the scattered field, represented by the vector $\mathbf{p}$, is expressed by the matrix-vector product:
\begin{equation}
    \mathbf{p} =  \mathbf{T}  \mathbf{a}\, .
\end{equation}
Here, the T-matrix $\mathbf{T}$ appears, which contains all the information about how an object interacts with light in linear approximation. After acquiring the T-matrix, one can compute the response for an arbitrary linear incident field. This makes it a very well-suited object for storage purposes. 

For spheres, the T-matrix is computed analytically via Mie theory~\cite{mie1908beitrage}. For arbitrary shapes, numerical solvers of Maxwell's equations must be employed~\cite{demesy_scattering_2018, fruhnert2017computing,hohenester_nanophotonic_2022,  ganesh2010three}. Since computing the T-matrix with a full-wave solver up to a large enough multipole order $l_\mathrm{max}$ for a non-axisymmetric scatterer includes at least $2 l_\mathrm{max} (l_\mathrm{max}+2)$ illumination scenarios, this results in a substantial computational effort that is inefficient to perform for just a single subsequent use of the T-matrix. In order to ensure consistency between different computation codes, we accept a fixed normalization convention for the VSWs, following Jackson~\cite{jackson1998classical} (different from~\cite{doicu2006light}). The ordering of the basis modes is also fixed as described in~\cite{ASADOVA2025109310}. 

The T-matrix-based scattering approach is particularly beneficial when a scatterer will be used in many different arrangements~\cite{foldy1945multiple, Mackowski:96}. That is because the computational effort lies in retrieving the T-matrix for each scatterer, but the effort required to compute the response of more complex photonic materials composed of these scatterers is rather low. Translation theorems for VSWs allow for expanding the field scattered by one scatterer at the origin of another scatterer. This allows the scattered field from one scatterer to be included in the incident field on all the other scatterers, enabling a full multiscattering description within the same algebraic framework. Generally, the multi-scattering problem can be expressed as
\begin{equation}\label{toglobal}
 \mathbf{p}= \left(\mathbb{1} - \mathbf{T} \mathbf{C}^{(3)}\right)^{-1} \mathbf{T} \mathbf{a}
\, , 
\end{equation}
where $\mathbf{T}$ is a block-diagonal matrix made of individual T-matrices of the scatterers, $\mathbf{C}^{(3)}$ is the matrix with translation coefficients, and $\mathbb{1}$ is the identity matrix.
Not only can finite clusters be described by this concise formula, but also infinitely periodic structures, the response of which will come up in this work as well. For efficient evaluation of the periodic sums, we use the Ewald summation method~\cite{ewald1921berechnung}.  It is important to keep in mind that when the circumscribing spheres of individual scatterers overlap, the underlying spherical wave expansions are no longer guaranteed to be valid. However, several strategies can circumvent this limitation~\cite{bertrand_global_2020, Martin2019, lamprianidis2023transcending, schebarchov2019mind}.

\section*{T-matrix database}

\begin{figure}[!t]
  \centering
  \includegraphics[width=0.8\linewidth]{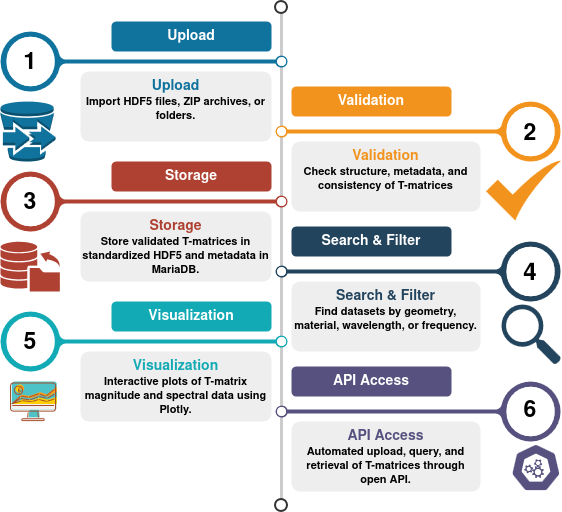}
 \caption{Overview of the main features of the Daphona T-matrix Portal, including upload, validation, storage, search and filtering, visualization, and API access.}
\label{fig:Workflow}
\end{figure}

To support structured access, exploration, and reuse of T-matrix datasets, we developed the Daphona T-matrix Portal, a web-based platform for interactive searching, filtering, and exporting standardized light-scattering data in terms of T-matrices. It serves as the primary user interface to the core T-matrix database, supporting research workflows built around precomputed light-scattering data. Rather than treating T-matrices as simple uploaded files, the portal imports them as structured scientific records with metadata describing the scatterer geometry, material properties, spectral range, and computation details. Before adding a T-matrix file to the database, it goes through a validation step that ensures two key points: (i) the dataset is not already stored, and (ii) it fully follows the predefined T-matrix data standard introduced in our related publication \cite{ASADOVA2025109310}. Only validated and compliant datasets are indexed and searchable. This approach allows users to access consistent, interpretable, and reliable data for further simulations and analysis.

The portal provides a structured filtering system that allows users to extract physically consistent subsets of T-matrix data. The results can be refined based on geometry, material type, spectral range, and other metadata, enabling precise and application-oriented dataset selection. Filters are applied dynamically, allowing users to refine queries step by step without interrupting their workflow. Consequently, the Daphona Portal facilitates both precise data retrieval and comprehensive exploration within large T-matrix databases.

\subsection*{System Overview and Functional Workflow}

The Daphona T-matrix Portal is designed as a complete infrastructure for the ingestion, validation, storage, discovery, and reuse of precomputed T-matrices. Figure~\ref{fig:Workflow} illustrates the functional workflow. Users upload datasets either as individual HDF5 files, compressed archives, or complete directory collections. Each submission undergoes automated validation to ensure structural correctness, metadata completeness, and compliance with the T-matrix data standard \cite{ASADOVA2025109310}. Only compliant datasets are accepted into the repository and stored in the standardized HDF5 format, along with structured metadata, in a relational database.

Once stored, datasets become accessible through an interactive web interface that supports physics-aware search and filtering, as well as visualization of selected scattering quantities. In addition to the graphical interface, the platform exposes an open API to support automated upload, querying, and retrieval workflows, enabling integration into simulation pipelines and reproducible research environments.

\begin{figure}[!t]
  \centering
  \includegraphics[width=\linewidth]{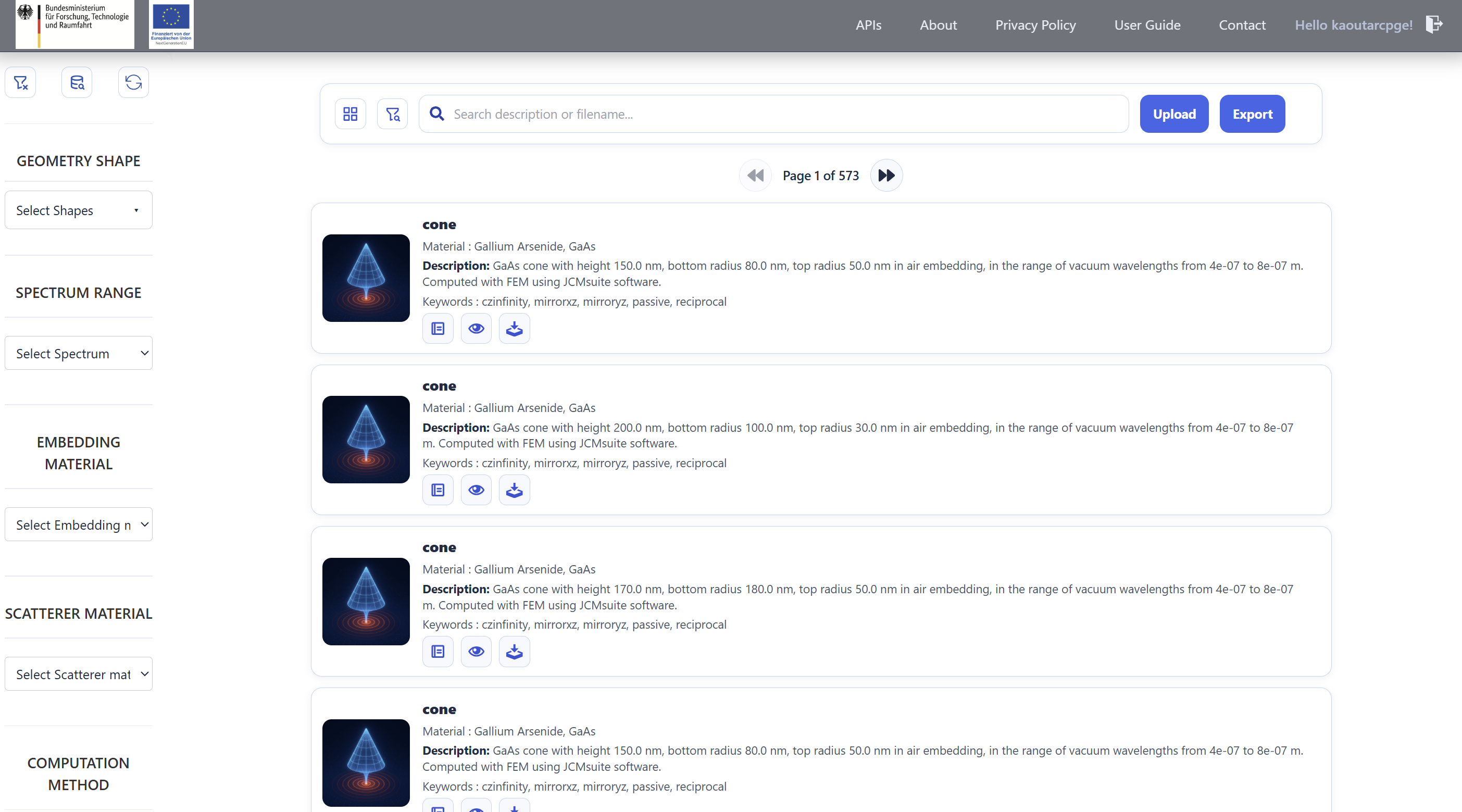}
 \caption{Main exploration interface of the Daphona T-matrix Portal, showing the searchable repository view with structured dataset cards and physics-aware filtering controls.}
\label{fig:MainPage}
\end{figure}
\subsection*{Dataset Repository and Exploration Interface}

The main interface of the Daphona T-matrix Portal provides a structured and intuitive workspace for exploring large collections of precomputed T-matrices. After authentication, users are presented with a centralized repository view containing hundreds of scientifically curated datasets. Each entry is displayed as an interactive card summarizing the essential physical and computational metadata, including geometry, material, dimensional parameters, spectral range, and associated keywords. Pagination, grid/list visualization modes, and a global search bar support efficient navigation.

The left-hand sidebar provides physics-aware filtering controls that allow users to constrain the displayed datasets by geometry type, spectral range, embedding medium, scatterer material, or numerical computation method. In this way, the main page is not only a browsing interface but also a scientific workspace that enables users to rapidly identify and select relevant T-matrix datasets for their applications.

\subsection*{T-matrix file, Validation, and Programmatic Access}

\begin{figure}[t]
\centering

\begin{subfigure}[t]{0.48\textwidth}
    \centering
    \includegraphics[width=\linewidth]{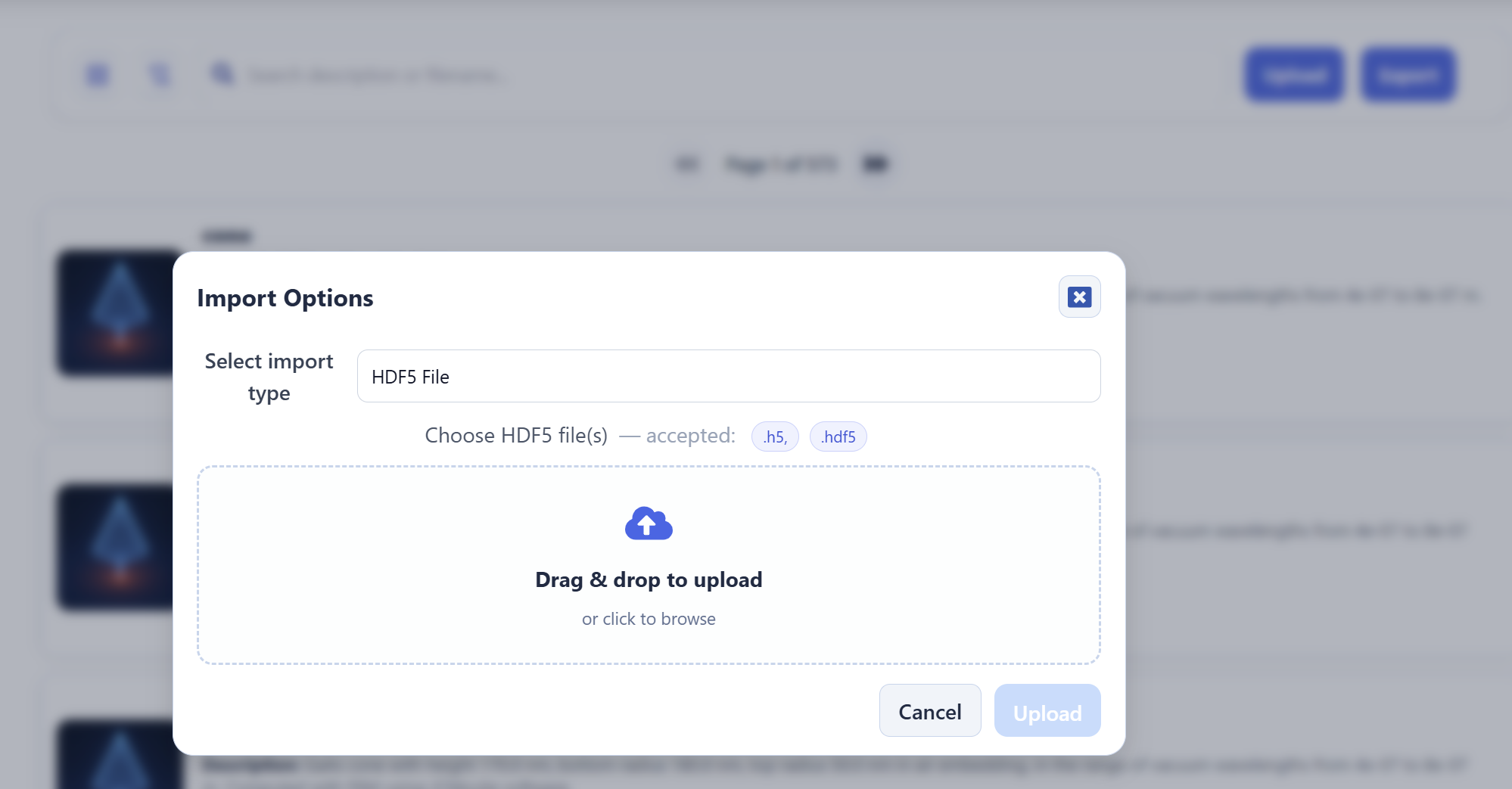}
    \caption{Upload via user interface.}
\end{subfigure}
\hfill
\begin{subfigure}[t]{0.48\textwidth}
    \centering
    \includegraphics[width=\linewidth]{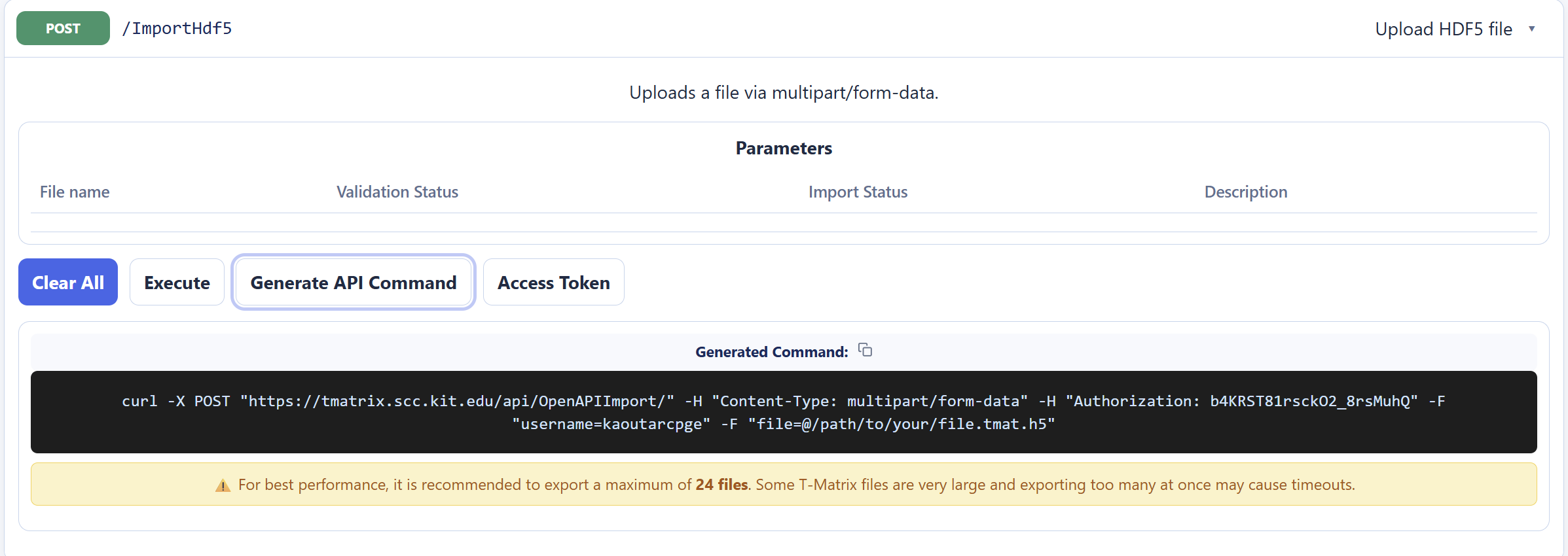}
    \caption{Upload via OpenAPI interface.}
\end{subfigure}

\begin{subfigure}[t]{0.48\textwidth}
    \centering
    \includegraphics[width=\linewidth]{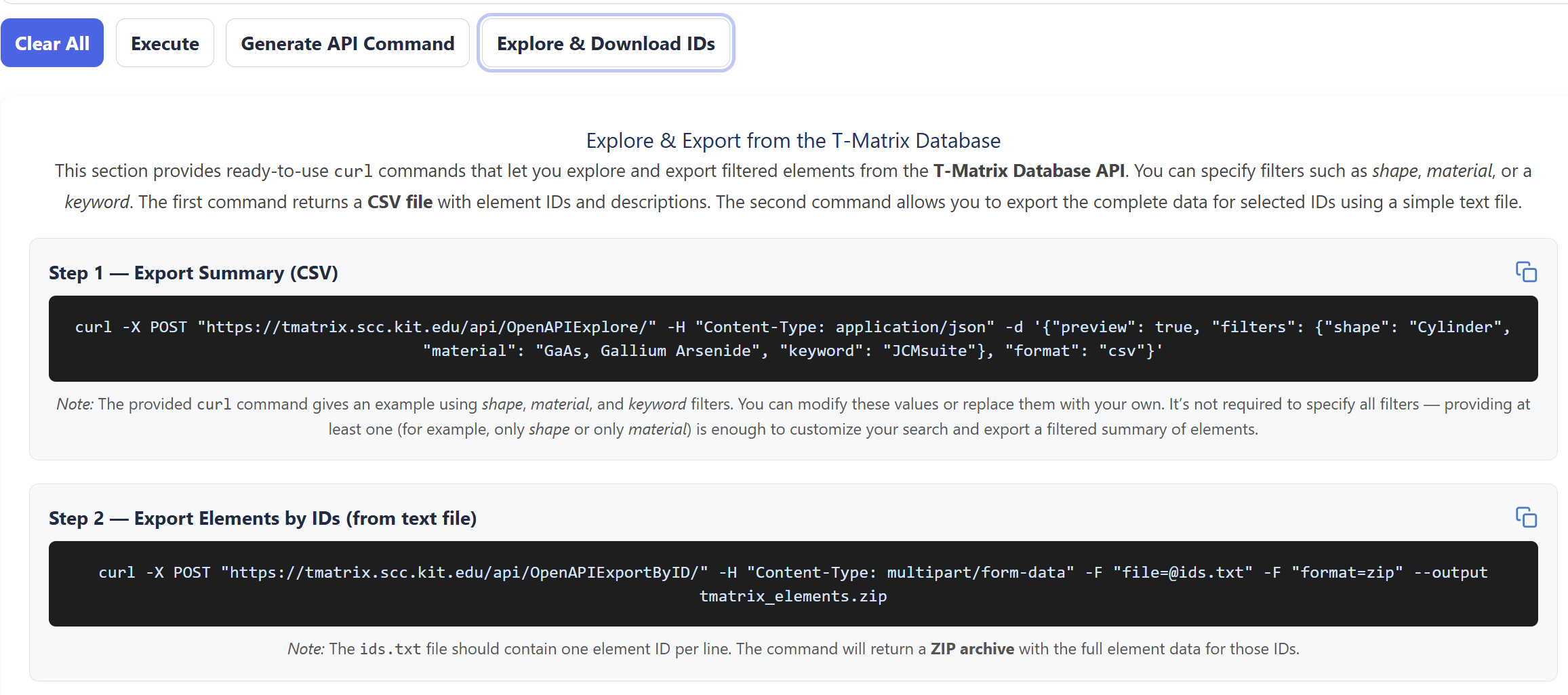}
    \caption{Explore and download filtered datasets via API.}
\end{subfigure}
\hfill
\begin{subfigure}[t]{0.48\textwidth}
    \centering
    \includegraphics[width=\linewidth]{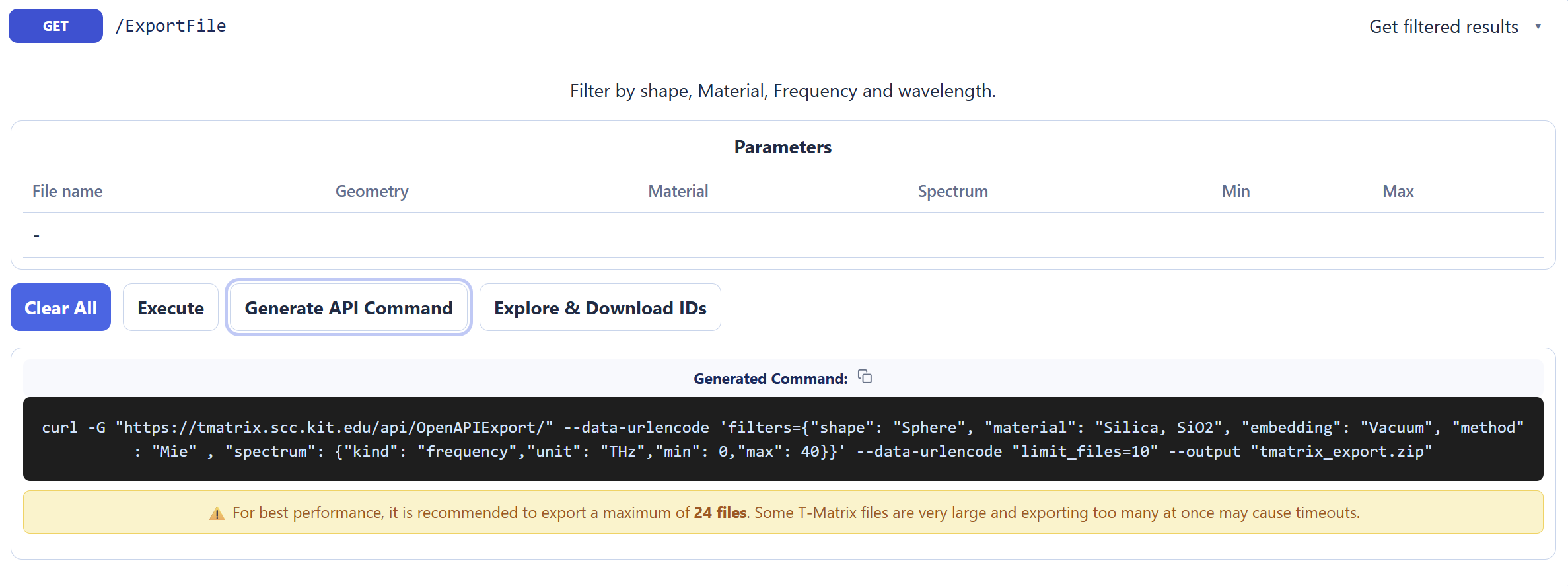}
    \caption{Export filtered datasets via API or generated commands.}
\end{subfigure}

\caption{Upload, exploration, and export capabilities offered by the Daphona T-matrix Portal through both the graphical user interface and the OpenAPI endpoints.}
\label{fig:upload_api_overview}
\end{figure}

The portal supports structured data upload through both an intuitive web interface and a programmatic OpenAPI endpoint (see Fig.~\ref{fig:upload_api_overview}). Users can upload individual HDF5 files, compressed archives, or complete directory collections of T-matrix datasets. Every submission is processed through a dedicated automated validation pipeline that rigorously examines structural integrity, metadata completeness, internal physical consistency, and full conformance with the published T-matrix data standard \cite{ASADOVA2025109310}. The system also performs duplicate detection to avoid redundant storage and ensure clean dataset provenance.

If a dataset does not meet the required criteria, the upload is safely rejected, and the user is provided with detailed diagnostic feedback, including structured error messages and guidance on how to correct the file. Validated datasets, in contrast, are admitted into the repository, indexed, and immediately integrated into the searchable database.
Once accepted, datasets can be retrieved either interactively through the graphical interface or programmatically via the OpenAPI endpoints. The export functionality supports both filtered and bulk retrieval, enabling seamless integration into numerical simulation workflows and large-scale optical studies. Through this combination of robust upload, strict validation, transparent feedback, and flexible export capabilities, the Daphona T-matrix Portal operates not merely as a browsing platform but as a trusted scientific infrastructure for high-quality, standardized, and reusable electromagnetic scattering data. The data it contains can be used to address specific scientific questions in the broader context of information-driven research. We will elaborate on that in the following section.

\section*{Machine learning with T-matrix data}
% There has been extensive research into machine learning applications for the problems commonly encountered by the nanophotonics community. There are typically two kinds of problems of interest. Predicting the optical response for a given scattering structure is the forward problem\cite{wiecha2019deep, augenstein2023neural, rezaei_faster_2024}, and, conversely, predicting the structure that produces a desired optical response is considered the inverse problem~\cite{so2020deep, wiecha2021deep, ma2025benchmarking}. 
As introduced above, deep learning approaches in nanophotonics address both forward and inverse tasks. Here, we focus on how T-matrix data supports these tasks.
The forward problem is deterministic and can be computed with dedicated software. Nevertheless, a trained model can greatly increase prediction speed. The inverse problem is ill-posed and non-unique at its core, so the inverse mapping cannot be obtained by analytic inversion. 

The clear advantage of using a T-matrix in these workflows is that the compactly stored data contains the response for illuminations with all possible polarizations and angles of incidence when plane waves are considered. This enables efficient generation of training data tailored to various objectives. While the stored T-matrices typically correspond to a single scatterer, widely used multiscattering codes allow us to simulate the optical response of periodic and non-periodic arrangements of the given scatterers and use their response as objectives as well~\cite{mackowski2011multiple,egel2017celes, egel2021smuthi, nevcada2021multiple, schebarchov2022multiple, shalev2024multem, beutel2024treams}. The existence of a T-matrix database, therefore, provides a convenient, ready-made source of training data.

To demonstrate the practical utility of the database, we present a few illustrative examples and structure the presentation of the results into forward and inverse problems, respectively. The examples rely on a subset of the data contained in the database, chosen for illustrative purposes. For that, we retrieve from the database the T-matrices from three geometric shapes: cylinders, cones, and rectangular cuboids, which are all made of silicon and embedded in air. The corresponding T-matrices are available over a wavelength range from 700~nm to 1000~nm, and that is the spectral range we consider in our analysis. Each T-matrix is truncated at the fifth multipole order. The resulting dataset comprises approximately 3000 samples per class, which is compact yet suitable for our purposes. This dataset is used in all of the following examples. 
\subsection*{Solving the forward problem}
We first address the forward problem and predict the optical response from a given set of parameters characterizing the scatterers. Here, we consider two routes. One can either predict the T-matrix of a single scatterer on its own or predict the optical response of the relevant actual arrangement of scatterers, since these properties can be computed quickly with available multiscattering codes. A T-matrix is typically higher-dimensional than, for example, a reflectance spectrum from a metasurface, and, therefore, seems to be more difficult to predict. We would like to investigate whether comparable prediction accuracy can be achieved in these two cases. 
\subsubsection*{Predicting the T-matrix}
First, we set the objective to be the T-matrix, represented as a concatenation of the real and imaginary parts of its individual entries. The data for each wavelength are treated as separate input samples. When splitting the data into training, validation, and test sets, we ensure that geometries present in the training set do not appear at different wavelengths in the validation and test sets. 

\begin{figure}[!htp]
  \centering
  \includegraphics[width=0.8\linewidth]{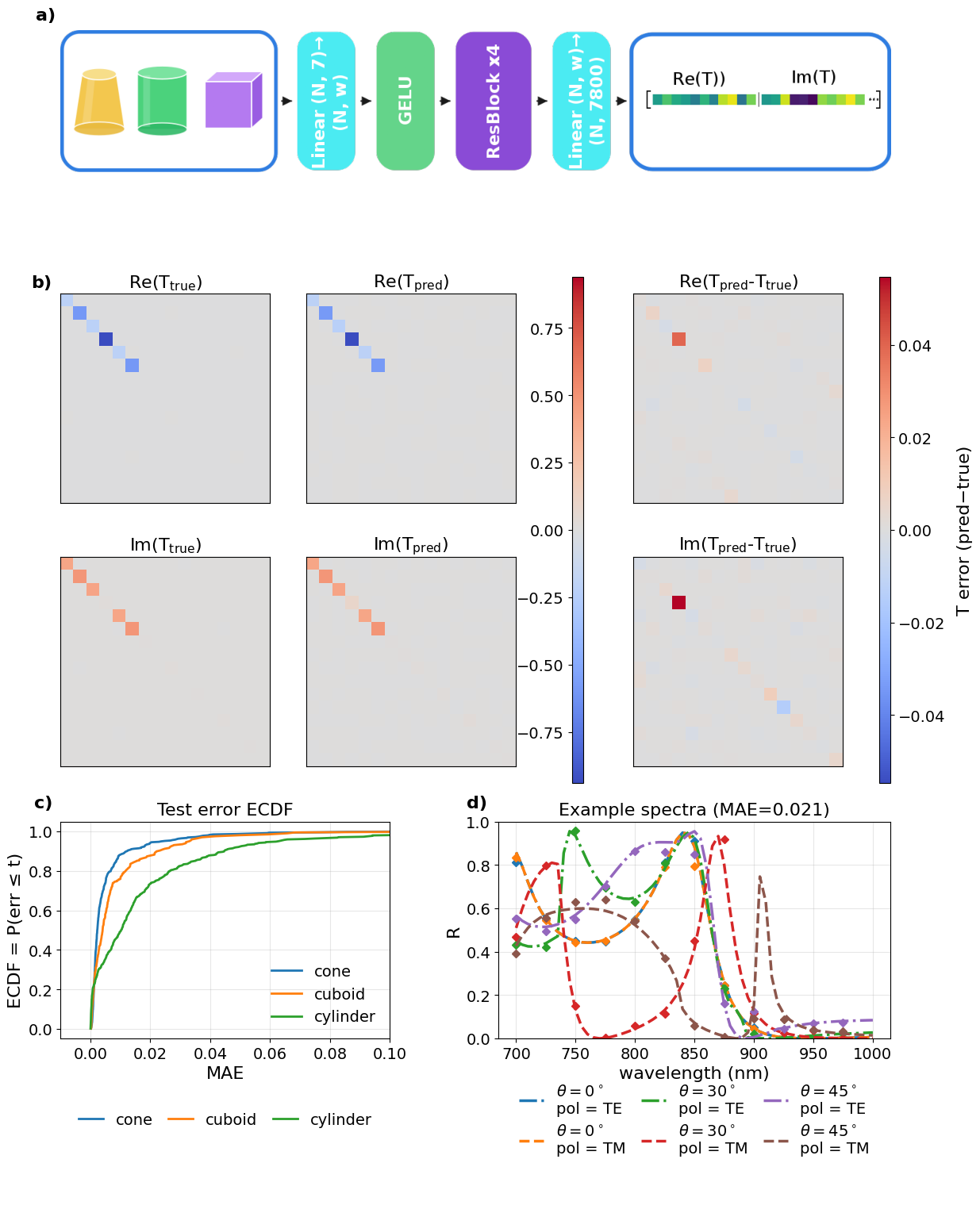} % or .png
  \caption{Test set evaluation of the surrogate model predicting T-matrices.
a) Schematic of the surrogate network: The model maps geometry class and parameters to the real and imaginary parts of the $\mathrm{T}$-matrix of individual scatterers.
% (b) The empirical cumulative distribution function (ECDF) of the symmetric relative Frobenius loss in the predicted $\mathrm{T}$-matrix per geometry class of those samples contained in the test data.
% (c) Real and imaginary parts of the $\mathrm{T}$-matrix elements for the reference ($\mathrm{T}_{\mathrm{true}}$) and the prediction ($\mathrm{T}_{\mathrm{pred}}$), together with the residual $\mathrm{T}_{\mathrm{pred}}-\mathrm{T}_{\mathrm{true}}$ for a representative test geometry: a cylinder with radius 130\,nm and height 170\,nm.
% (d) Predicted (markers) and reference (solid lines) reflectance spectra $R(\lambda)$ for the same sample at three angles of incidence and two polarizations. 
(b) Real and imaginary parts of the $\mathrm{T}$-matrix elements for the reference ($\mathrm{T}_{\mathrm{true}}$) and the prediction ($\mathrm{T}_{\mathrm{pred}}$), together with the residual $\mathrm{T}_{\mathrm{pred}}-\mathrm{T}_{\mathrm{true}}$ for a representative test geometry: a cylinder with radius 110\,nm and height 190\,nm.
(c) The empirical cumulative distribution function (ECDF) of the reflectance values computed from the predicted T-matrices per geometry class of those samples contained in the test data.
(d) Predicted (markers) and reference (solid lines) reflectance spectra $R(\lambda)$ for the same sample at three angles of incidence and two polarizations. 
}
\label{fig:ptot}
\end{figure}

The neural network that we consider here is illustrated in Fig.~\ref{fig:ptot}(a). As an input, it takes from each dataset the class label, the normalized geometrical parameters, and the normalized wavelength, and passes this through a multi-layer perceptron (MLP) consisting of an initial fully connected layer with a GELU activation, followed by four residual blocks of width 
1024. The final layer spills out the vector containing the T-matrix of the given object. In our implementation, we use PyTorch for setting up the neural network and train with the AdamW gradient-based optimizer. This also holds for all the models considered below. The network has a total of 27,021,528 trainable parameters. The hyperparameters were selected through a manual sweep based on validation performance. The entire network, as illustrated in Fig.~\ref{fig:ptot}(a), eventually serves as a surrogate solver for Maxwell's equations that predicts the T-matrix for a given input object.

We train the network using a normalized symmetric Frobenius loss on the T-matrix, defined as
\begin{equation}
    \ell = \frac{\lVert \mathrm{T}_{\text{pred}} -  \mathrm{T}_{\text{true}} \rVert_F^2}
               {\lVert  \mathrm{T}_{\text{pred}} \rVert_F^2 + \lVert  \mathrm{T}_{\text{true}} \rVert_F^2 + \varepsilon}\, ,
\end{equation}
where $\varepsilon>0$ is a small constant introduced for numerical stability.
Compared to a standard mean-squared error on the T-matrix entries, this relative normalization yields a more balanced weighting across the dataset.
Models were trained for 150 iterations, during which convergence was reached.

It should be noted that an overall judgment on how useful a predicted T-matrix is is difficult to provide, as the individual entry does not necessarily correspond to any of the experimentally observable quantities typically considered. Therefore, we also study the reflectance of a metasurface made from a periodic arrangement of the considered scatterer. For that purpose, we compute the reflectance of an infinitely extended two-dimensional periodic arrangement of the scatterer using the predicted T-matrix as input to the multiscattering code \texttt{treams}\cite{beutel2024treams}. 
The lattice period is set based on the diameter of the circumscribing sphere of the scatterer, with an additional $100\,\mathrm{nm}$ gap in both in-plane directions to define the unit cell. Reflectances are calculated for plane wave illumination at three angles of incidence, $0^{\circ}, 30^{\circ}$, and $45^{\circ}$, and two orthogonal polarizations, TE and TM, in parity basis. This holds throughout the further examples as well. 

The results are evaluated on a test set and depicted in Figs.~\ref{fig:ptot}(b)-(d). 
To visualize a T-matrix that we can predict with our surrogate solver, we show a representative test sample in Fig.~\ref{fig:ptot}(b). We select the test sample whose reflectance spectra exhibit the largest level of variation, and display its predicted and reference T-matrices at the wavelength with the largest prediction error for that geometry. In that sense, the sample can be considered the most difficult sample among all the samples contained in the test set. The specific geometry of the considered object is a cylinder  with radius
110 nm and height 190 nm. The figure shows the error between the predicted T-matrix and the ground truth as well. We can clearly see that, even in this example, the overall T-matrix is fairly accurately predicted, while only the individual components show errors. 

The empirical cumulative distribution function (ECDF) of the test error, which is the fraction of test samples with mean absolute error (MAE) in R below a threshold $t$, reveals class-dependent accuracy as shown in Fig.~\ref{fig:ptot}(c). Cones exhibit the lowest errors (geometry-fair median $0.002$ and $90$th percentile $0.012$), followed by cuboids (median $0.004$, $p_{90}=0.022$), while cylinders are the most challenging (median $0.01$, $p_{90}=0.045$). This class dependence can be explained by differences in the underlying spectral dependence and in the parameter distribution of the dataset. The cylinders exhibit sharper resonances in the considered wavelength range, whereas cones occupy a more compact region in parameter space, making prediction easier. 

The predicted and reference reflectance spectra in  Fig.~\ref{fig:ptot}(d) reproduce the main spectral features across the considered wavelength range. The overall spectral response, as well as spectral details, are accurately predicted, even for the most difficult sample in the test set. The previously encountered error in the individual entry of the T-matrix does not significantly degrade our ability to predict this observable quantity.

\begin{figure}[!htp]
  \centering
  \includegraphics[width=\linewidth]{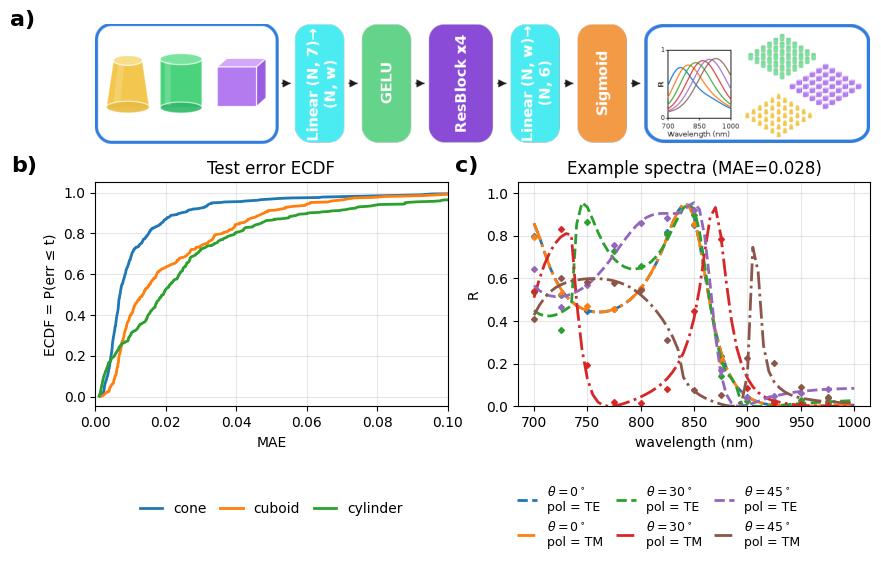} % or .png
  \caption{
  Test set evaluation of the reflectance surrogate model.
(a) Schematic of the surrogate network: the model maps geometry class and parameters to reflectance spectra over wavelength for multiple incidence angles and polarizations.
(b) ECDF of the mean-absolute error (MAE) of the predicted reflectance spectra, shown per geometry class.
(c) Predicted (markers) and reference (solid lines) reflectance spectra $R(\lambda)$ for a representative test geometry at angles of incidence $0^{\circ}$, $30^{\circ}$, and $45^{\circ}$ for TE and TM polarizations. Predictions are evaluated on a denser wavelength grid than the actual training grid. 
}
\label{fig:ptor}
\end{figure}
%   \caption{
%   Test set evaluation of the reflectance surrogate model.
% (a) CDF of the mean-absolute error (MAE) of the predicted reflectance spectra, shown per geometry class.
% (b) Predicted (markers) and reference (solid lines) reflectance spectra $R(\lambda)$ for a representative test geometry at angles  of incidence $0^{\circ}$, $30^{\circ}$, and $45^{\circ}$ for TE and TM polarizations.
% }
\subsubsection*{Predicting the reflectance spectra from a metasurface}
Next, we evaluate the direct prediction of reflectance spectra in a forward problem and train a dedicated network to predict reflectance rather than all entries of the T-matrix. The same residual MLP-backbone with the same inputs is used as before, but the head is simplified due to the lower-dimensional output. The network has a total of 8,411,142 trainable parameters in this case. In the end, 
a sigmoid activation layer is applied to constrain the prediction within the physical range. After training for the same number of iterations as the previous network, we evaluate the same characteristic curves. The ECDF curves shown in Fig.~\ref{fig:ptot}(b) are now computed with respect to the mean absolute error (MAE) of the reflectance spectra. In this case, the ECDF curves are rising faster while preserving the same ordering in performance.
% cones still show the best performance, while the cuboid error distribution shifts toward the cylinder one, indicating similar difficulty.

%%%%%NOT really happening when both are errors in R!
% In the T-matrix setting, known physical symmetries imposed strong structure on the target, which can make differences between classes more visible in the error distributions. When predicting reflectance directly, the measured spectra provide a more compressed representation, so the class-wise CDFs could look more similar.
The spectra of the same representative example in Fig.~\ref{fig:ptor}(c) reproduce well the overall spectral behavior. Using the current network, cones still come out best (geometry-fair median $0.007$ and $p_{90}=0.024$), cuboids follow (median $0.013$, $p_{90}=0.048$), and cylinders remain the hardest case (median $0.013$, $p_{90}=0.061$). Compared to these results, the previous network performs better, with lower errors both at the median and the 90$^\mathrm{th}$ percentile. This remains true even after bringing the parameter counts of the two networks closer to each other. This suggests that predicting the T-matrix as the output could effectively act as the first part of a physics-informed neural network, where the physical layer is implemented through multiscattering formalism.
Moreover, since predicting the T-matrix allows fast computation for any observed optical response for an arbitrary arrangement, it can be more beneficial to implement.
\subsection*{Solving the inverse problem}
Having established surrogate models to solve the forward problem, we next address inverse design. As in the previous section, we structure our work into two subsections. On the one hand, we show how to infer the geometrical parameters of a scatterer so that it provides a predefined T-matrix. On the other hand, we show how to infer the geometrical parameters of a scatterer so that it provides a predefined reflection spectrum when periodically arranged in space to form a metasurface.  

\subsection*{Prediction of the geometrical parameters from the T-matrix}
The first task we consider here is to infer the shape class and geometric parameters so that the scatterer will provide a target T-matrix. This problem appears in several settings, ranging from validation of declared metadata for $\mathrm{T}$-matrices submitted to the database, to inverse design scenarios where one seeks a geometry whose $\mathrm{T}$-matrix exhibits a prescribed structure, for example, specific relations between electric and magnetic multipole contributions.

\begin{figure}[!htp]
  \centering
  \includegraphics[width=\linewidth]{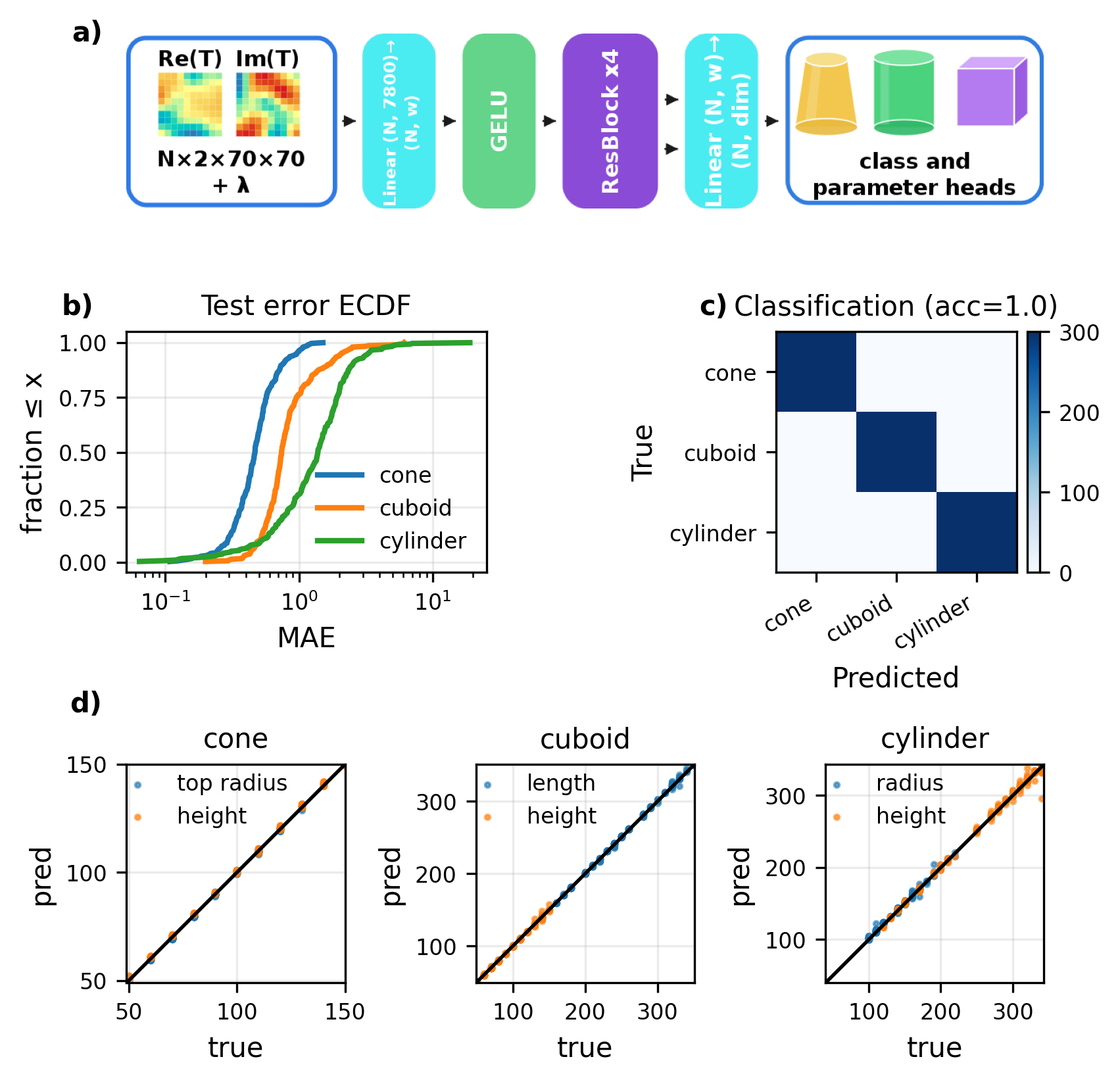} % or .png
 \caption{Test set evaluation of the multitask inverse model from T-matrix inputs.
 % (a) Schematic of the network: the T-matrix serves as input, a convolutional encoder extracts latent features, which are fed into separate heads for shape classification and geometry regression.
  (a) Schematic of the network: the T-matrix and wavelength serve as input, a residual MLP backbone produces latent features, which are fed into separate heads for shape classification and geometry regression.
(b) Per-class empirical cumulative distribution function (ECDF) of the per-sample regression error.
(c) Confusion matrix of predicted shape labels (cone, cuboid, cylinder).
(d) Parity plots of predicted versus ground-truth geometry parameters for cones, cuboids, and cylinders; the diagonal line indicates perfect agreement.}
\label{fig:ttop}
\end{figure}

We have to add that the three shapes can be distinguished already from the T-matrix due to symmetry considerations.
In particular, cuboids exhibit mirror symmetry about the $z$-axis, which can be verified through the following formula for the parity basis:
\[
T^{ij}_{\ell,m;\ell',m'} = (-1)^{\,i+j+\ell+\ell'+m+m'}\,T^{ij}_{\ell,m;\ell',m'}\, .
\]
Cones are axisymmetric and, therefore, satisfy the rotational-symmetry constraints
\[
T^{ij}_{\ell,m,\ell',m'}=\delta_{mm'}\,T^{ij}_{\ell,m,\ell',m'}\, ,
\qquad
T^{ij}_{\ell,m,\ell',m'} = (-1)^{i+j}\,T^{ij}_{\ell,-m,\ell',-m'}\, ,
\]
where the polarization indices $i,j\in\{1,2\}$.
Cylinders satisfy both of the above selection rules simultaneously. However, we consider classification an additional task to evaluate the performance of the network.

\begin{figure}[!htp]
  \centering
  \includegraphics[width=\linewidth]{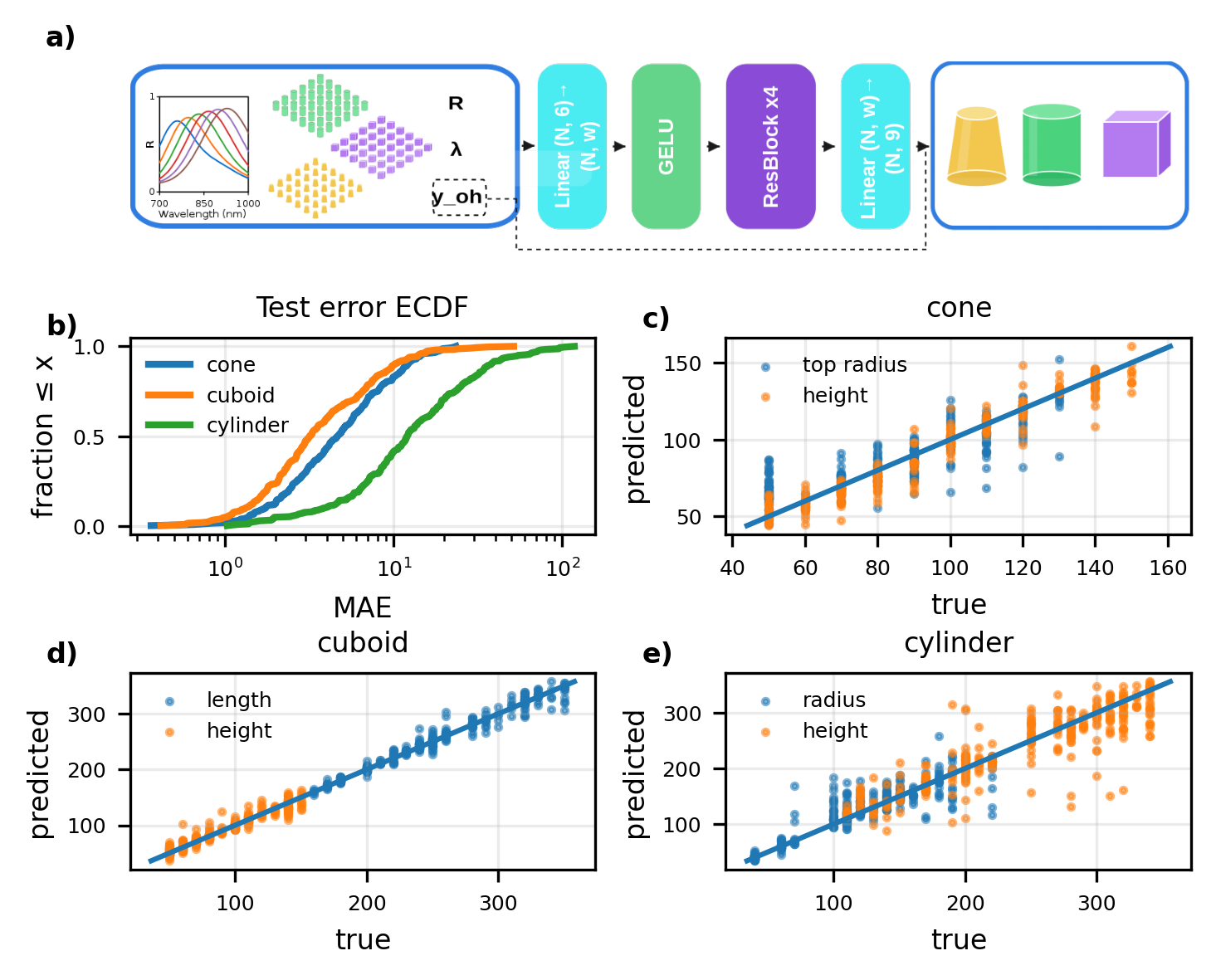} % or .png
 \caption{Test set evaluation of the multitask inverse model from reflectance inputs.
(a) Per-class empirical cumulative distribution function (ECDF) of the per-sample regression error.
(b) Parity plots comparing predicted vs.\ true geometric parameters for each class (shown for two representative parameters per shape); the diagonal denotes perfect agreement.}
\label{fig:rtop}
\end{figure}

There are different approaches to training the model. Here, we use the entire T-matrix directly as an input and predict the geometric parameters from it. Figure~\ref{fig:ttop}(a) provides an indication of the artificial neural network used for the purpose. 
% The network starts with a small convolutional encoder, chosen because the input naturally represents a structured two-dimensional array, and because convolutions can leverage the spatial relationships of multipole couplings. Next, adaptive average pooling compresses these features into a fixed-length latent vector, and dropout is applied for regularization. The wavelength is also embedded with an MLP and added to the pooled features. 
In addition to the normalized T-matrix entries, we also include the normalized wavelength as an input. Next, the network proceeds in the same way as the previous MLP-based models, using an initial fully connected layer followed by residual blocks.
The network is trained in a multitask setting. Therefore, one of the final heads predicts the shape class, and the other regresses the corresponding geometric parameters.
The total loss is a weighted sum
\begin{equation}
\mathcal{L} = w_{\mathrm{ce}}\,\mathcal{L}_{\mathrm{cls}} + w_{\mathrm{reg}}\,\mathcal{L}_{\mathrm{reg}}\, ,
\end{equation}
where $w_{\mathrm{reg}}$ is gradually increased during training. 
% We observe the better error distribution of the cone dataset in Fig.~\ref{fig:ttop}(b), and similarity between cuboid and cylinder error distributions.
As previously, we observe the better error distribution of the cone dataset in Fig.~\ref{fig:ttop}(b), followed by the cuboid and cylinder datasets.
The strong performance is evident in the perfect classification in Fig.~\ref{fig:ttop}(c), and the regression error is also small, as indicated by the tight clustering around the parity line in Fig.~\ref{fig:ttop}(d).  

\subsection*{Prediction of the geometrical parameters from the reflectance spectra of a metasurface}
As a complementary inverse task for the surrogate network, we consider a different setting in which the metasurface reflectance values are used as the network input. This is, in principle, an architecture-mirrored version of the forward surrogate with swapped input/output roles, as can be seen in Fig.~\ref{fig:rtop}(a), except that the shape label is given, and the sigmoid function is omitted.
We additionally condition the model on the wavelength and on the desired shape label to reduce ambiguities arising from the non-uniqueness of the inverse mapping. Regression is an even harder task in this setting because reflectance values are not uniquely informative, and similar values can arise from different geometries. This makes the problem intrinsically ill-posed. It can be seen that the model performance is limited, as evidenced by a wide spread in parity plots in Figs.~\ref{fig:rtop}(c)-(e), as well as much larger errors in the ECDF plot in Fig.~\ref{fig:rtop}(b). When using the full T-matrix, the inverse mapping is not guaranteed to be unique either. However, within the distinguishable families of shapes considered here, the ambiguity is reduced.

Overall, these examples show how the T-matrix database can serve as a starting point for both forward modeling and inverse design, with targets ranging from full T-matrices to measurable spectra. At the same time, once we move to compressed observables such as reflectance, the inverse problem can become even more ambiguous, underscoring the value of access to the full T-matrix.

\section*{Conclusion}
We have introduced the Daphona T-matrix Portal as a practical way to store, validate, and reuse T-matrix datasets in a consistent, standardized form. We invite the broader community to make use of it. By combining a unified HDF5 structure with automated validation and metadata indexing, the portal makes it straightforward to find relevant scattering data and to retrieve it either through the web interface or via an API for scripted workflows.

The machine-learning examples highlight why access to full T-matrices is useful beyond simple data archiving. For a forward problem, surrogate models can predict T-matrices with reasonable accuracy and, when these predictions are passed to a multiscattering solver, the resulting reflectance spectra reproduce the main trends and resonant features of the reference simulations. For an inverse problem, learning directly from the T-matrix enables reliable shape identification and accurate parameter regression for the considered shape families, whereas learning from reflectance alone remains less robust due to the inherent ambiguity of the objectives. 

Overall, the presented infrastructure demonstrates that standardized T-matrix datasets can be reused efficiently across different numerical and machine-learning-based workflows, thereby improving consistency and reproducibility and strengthening information-driven research.
We encourage the community to use the portal and to contribute additional datasets and feedback, with the goal of increasing dataset coverage and reducing redundant computations through reuse. With growing coverage and consistent metadata, the repository can also facilitate benchmarks of different T-matrix computation methods on common test cases.

% \begin{backmatter}

% Content in the funding section will be generated entirely from details submitted to Prism. Authors may add placeholder text in this section to assess length, but any text added to this section will be replaced during production and will display official funder names along with any grant numbers provided. If additional details about a funder are required, they may be added to the Acknowledgment, even if this duplicates some information in the funding section. For preprint submissions, please include funder names and grant numbers in the manuscript.

\section*{Acknowledgment}
N.A. acknowledges support from the Karlsruhe School of Optics and Photonics (KSOP).

\section*{Disclosures}
The authors declare no conflicts of interest.

% \bmsection{Data availability} Data underlying the results presented in this paper are available in Dataset 1, Ref. [8].

% \bmsection{Data availability} Data underlying the results presented in this paper are available in Ref. [8].

\section*{Data availability} 
The data underlying the results presented in this paper are available via the Daphona portal~\url{https://tmatrix.scc.kit.edu/}. The scripts used to reproduce the results are available at \url{https://github.com/tfp-photonics/tdat_ml}.
%Data underlying the results presented in this paper are not publicly available at this time but may be obtained from the authors upon reasonable request.

% \noindent Data availability statements are not required for preprint submissions.

% \end{backmatter}

%%%%%%%%%%%%%%%%%%%%%%% References %%%%%%%%%%%%%%%%%%%%%%%%%

%%%%%%%%%% If using BibTeX:
\bibliography{sample}

%%%%%%%%%% If preparing manually:
% \begin{thebibliography}{1}
% \newcommand{\enquote}[1]{``#1''}

% \bibitem{Zhang:14}
% Y.~Zhang, S.~Qiao, L.~Sun, Q.~W. Shi, W.~Huang, L.~Li, and Z.~Yang,
%   \enquote{Photoinduced active terahertz metamaterials with nanostructured
%   vanadium dioxide film deposited by sol-gel method,}
%   {\protect\JournalTitle{Optics Express}} \textbf{22}, 11070--11078 (2014).

% \bibitem{Optica}
% {Optica}, \enquote{{Optica Publishing Group},}
%   \url{http://www.opg.optica.org}.

% \bibitem{FORSTER2007}
% P.~Forster, V.~Ramaswamy, P.~Artaxo, T.~Bernsten, R.~Betts, D.~Fahey,
%   J.~Haywood, J.~Lean, D.~Lowe, G.~Myhre, J.~Nganga, R.~Prinn, G.~Raga,
%   M.~Schulz, and R.~V. Dorland, \enquote{Changes in atmospheric consituents and
%   in radiative forcing,} in \enquote{Climate Change 2007: The Physical Science
%   Basis. Contribution of Working Group 1 to the Fourth Assesment Report of
%   Intergovernmental Panel on Climate Change,}  S.~Solomon, D.~Qin, M.~Manning,
%   Z.~Chen, M.~Marquis, K.~B. Averyt, M.~Tignor, and H.~L. Miler, eds.
%   (Cambridge University Press, 2007).

% \end{thebibliography}

\end{document}